# An Improvement in Quantum Fourier Transform

Laszlo Gyongyosi[*], Sandor Imre

*Quantum Technologies Laboratory*
*Department of Telecommunications*
*Budapest University of Technology and Economics*
*2 Magyar tudosok krt, Budapest, H-1117, Hungary*
[*]*gyongyosi@hit.bme.hu*
(Dated: 2010)

**Singular Value Decomposition (SVD) is one of the most useful techniques for analyzing data in linear algebra. SVD decomposes a rectangular real or complex matrix into two orthogonal matrices and one diagonal matrix. In this work we introduce a new approach to improve the preciseness of the standard Quantum Fourier Transform. The presented Quantum-SVD algorithm is based on the singular value decomposition mechanism. While the complexity of the proposed scheme is the same as the standard Quantum Fourier Transform, the precision of the Quantum-SVD approach is some orders higher. The Quantum-SVD approach also exploits the benefits of quantum searching.**

## 1 Introduction

Quantum Fourier Transform (QFT) is a fundamental transformation in quantum information processing[6-9,14-16,18-26]. QFT transformation samples the states of the input quantum register at equally-spaced angles. Non-uniform QFT transformation samples the input states of the quantum register at *non-equally*-spaced angles, and as a result, the QFT coefficients can also contain *unevenly* sampled angles. QFT transformation, using non-uniformly spaced or unequal angle intervals is equivalent to the classical QFT transformation, without the limitations of equally spaced angle intervals. Non-uniform QFT transformation can be applied to compute the equally spaced QFT angle intervals.

We propose a Quantum-SVD based approach for the computation of Quantum Fourier Transformation of non-uniform angles. The proposed Quantum algorithm interpolates the non-uniform angles of non-uniformly distributed states in the Fourier domain. In the proposed Quantum-SVD approach, first the non-uniform angles are sampled and the singular value decomposition (SVD) of the sampled angles is then computed. The left singular vectors corresponding to the singular values that are not close to zero are used to interpolate the non-uniform angles at all intervals. The quantum states are not sampled at equally spaced intervals; hence the non-uniform angles are not necessarily on the unit intervals.

## 1.1 Motivation

The uniformly sampled angles of ordinary QFT can be computed from unevenly sampled angles. The non-uniform QFT is the same as the classical QFT transformation, however in this case the quantum register is sampled at unequally spaced, irregular angle intervals. The non-uniform QFT first applies an ordinary QFT transformation on the input quantum register to get the ordinary QFT coefficients, on regularly spaced angle intervals. Using the ordinary QFT coefficients with regularly spaced angle intervals, the angle-space representation of the state of the input quantum register can be interpolated to find the values at the desired angle intervals. The interpolation only produces an approximation of the coefficients at these angle intervals, the precision of the approximation depends on the length of the input quantum register, and the spacing of the angle intervals[1,2]. The non-uniform QFT angle in the interval between two uniform QFT angles is approximated by a set of orthogonal vectors, using singular value decomposition.

In Fig. 1.a. we illustrated the uniform equally spaced angle intervals of ordinary QFT, and the interval between two uniform QFT angles.

We show the vectors of $e^{i2\pi\frac{xy}{N}}$ for $y = 0,...,N-1$. In Fig. 1.b. the ordinary, equally spaced QFT angles are denoted by grey dots, the non-uniform QFT angle in the interval between two uniform QFT angles is denoted by the grey area with black dot and dashed arrow. In our Quantum-SVD approach, the non-uniform QFT angle in the interval between two uniform QFT angles will be approximated by a set of orthogonal vectors.

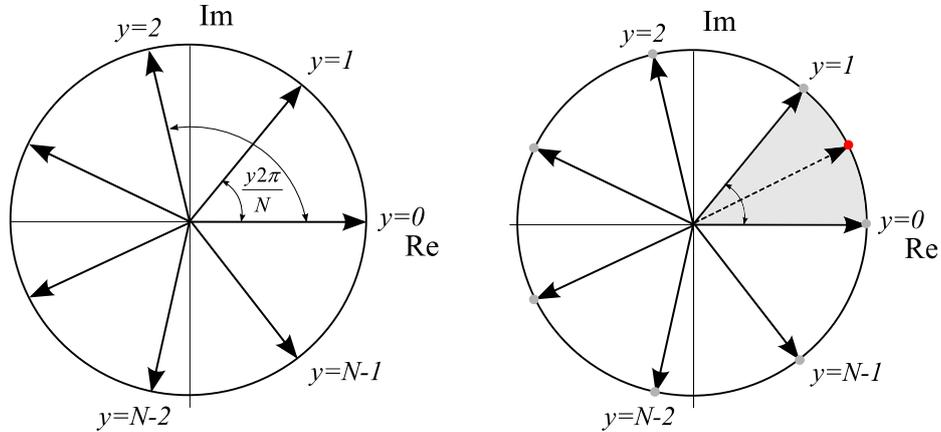

**Fig. 1.** Uniform equally spaced angle intervals of ordinary QFT (a), and a non-uniform QFT angle in the interval between two uniform QFT angles (b).

The Quantum-SVD algorithm is a novel approach for the computation of the Quantum Fourier Transformation of non-uniformly distributed states. In the proposed Quantum-SVD algorithm, the ordinary QFT transformation is applied to the scale input quantum register, and then the state of input quantum register is scaled according to a small set of vectors. In the final step, the QFT coefficients referring to the same angle are combined.

This paper is organized as follows. In Sections II and III, we discuss the basic properties of quantum information processing and QFT, plus the main elements of the singular value decomposition based approximation set. In Section IV, we present the proposed Quantum-SVD algorithm. In Section V, we analyze the precision of the algorithm, and compare the results with classical QFT. Finally, in Section VI, we conclude the results.

### 1.2 Problem statement

Before we start to discuss the proposed Quantum-SVD approach, we give the classical analogy to Quantum-SVD algorithm by a brief overview. Using the classical analogy, in many classical applications the data is sampled on a non-uniform grid, preferably non-uniformly spaced points in the frequency domain. In these cases, we can use a generalization of the DFT known as the non-uniform DFT. The non-uniform DFT transformation is very useful in applications, where input samples must be taken at irregular intervals[1,2].

The non-uniform DFT transformation allows for more selectively concentrated information. The original non-uniform DFT transformation is equal to an interpolation, which can be viewed as two serial transformations. In the first one, the input data is used to perform a uniform DFT, as a result we get Fourier coefficients on a regularly spaced grid. In the second step, the non-uniform DFT transformation uses the computed DFT coefficients, to interpolate the frequency representation of the original signal, at the desired frequency nodes[1,2].

In classical non-uniform DFT, for a given discrete signal $x_n$, $n = 0,...,N-1$, the result of the non-uniform DFT transform can be expressed as

$$X(k) = \sum_{n=0}^{N-1} x_n e^{-j2\pi(n-\tau)\frac{f_k}{N}}, \quad \text{where} \quad f_k \in [0, N) \quad \text{for} \quad k = 0,...,K-1, \quad \text{and}$$

$n = 0,...,N-1$. In this representation, $K$ is the cardinality of the set of non-uniform frequencies $f_k$, and $\tau$ is the shift parameter[1,2]. The shift parameter changes the indexing of DFT exponentials. For uniform DTF, $f_k$ includes uniform frequencies $0,...,N-1$. The shift parameter generally can be chosen to[1,2] $\tau = N/2$, and the input signal $X(k)$ is indexed in interval $-N/2,...,N/2-1$.

In conclusion, the non-uniform DFT receives frequency information at uniformly spaced frequency nodes and the results used to interpolate the desired frequency nodes. In the interpolation-phase, the non-uniform DFT only produces an approximation of the DFT coefficients at desired frequency nodes[1]. The precision of that approximation will depend on the specific behavior of the function being transformed and the spacing of the frequency nodes.

**1.3   Brief overview of quantum information processing**

In this section, we give a brief overview of quantum mechanics, and we introduce the basic notations which will be used in the text. In quantum information processing, the logical values of classical bits are replaced by state vectors $|0\rangle$ and $|1\rangle$, which notation is called Dirac notation. Contrary to classical bits, a qubit $|\psi\rangle$ can also be in a linear combination of basis vectors $|0\rangle$ and $|1\rangle$. The state of a qubit can be expressed as $|\psi\rangle = \alpha|0\rangle + \beta|1\rangle$, where $\alpha$ and $\beta$ are complex numbers, which is also called the superposition of the basis vectors, with probability amplitudes $\alpha$ and $\beta$. A qubit $|\psi\rangle$ is a vector in a two-dimensional complex space, where the basis vectors $|0\rangle$ and $|1\rangle$ form an orthonormal basis. The orthonormal basis $\{|0\rangle, |1\rangle\}$ forms the computational basis, in Fig. 2. we illustrate the computational basis for the case probability amplitudes real[7].

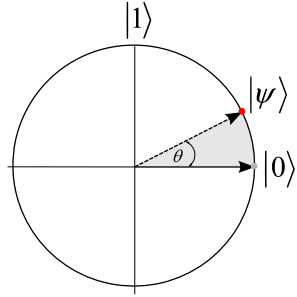

**Fig. 2.** Computational basis and general representation of a qubit in superposition state.

The vectors or states $|0\rangle$ and $|1\rangle$ can be expressed in matrix representation by

$$|0\rangle = \begin{bmatrix} 1 \\ 0 \end{bmatrix} \text{ and } |1\rangle = \begin{bmatrix} 0 \\ 1 \end{bmatrix}. \tag{1}$$

If $|\alpha|^2$ and $|\beta|^2$ are the probabilities, and the number of possible outputs is only two, then for $|\psi\rangle = \alpha|0\rangle + \beta|1\rangle$ we have $|\alpha|^2 + |\beta|^2 = 1$, and the norm of $|\psi\rangle$ is $\||\psi\rangle\| = \sqrt{|\alpha|^2 + |\beta|^2} = 1$. The most general transformation that keeps constraint of $\||\psi\rangle\|$ is a linear transformation $U$, that takes unit vectors into unit vectors. A *unitary* transformation can be defined as $U^\dagger U = UU^\dagger = I$, where $U^\dagger = (U^*)^T$, hence the adjoint is equal to the transpose of complex conjugate, and $I$ is the identity matrix[7].

The tensor product will have an important role in our Quantum-SVD algorithm, thus we shortly introduce the concept of tensor product. If we have complex vector spaces $V$ and $W$ of dimensions $m$ and $n$, then the tensor product of $V \otimes W$ is a $mn$ dimensional vector space.

The elements of $V \otimes W$ are linear combinations of tensor products $|v\rangle \otimes |w\rangle$, and for $x \in \mathbb{C}$, and $|v_1\rangle, |v_2\rangle \in V$ and $|w_1\rangle, |w_2\rangle \in W$ we obtain, that

$$x(|v\rangle \otimes |w\rangle) = (x|v\rangle) \otimes |w\rangle = |v\rangle \otimes (x|w\rangle), \tag{2}$$

$$(|v_1\rangle + |v_2\rangle) \otimes |w\rangle = (|v_1\rangle \otimes |w\rangle) + (|v_2\rangle \otimes |w\rangle), \tag{3}$$

and

$$|v\rangle \otimes (|w_1\rangle + |w_2\rangle) = (|v\rangle \otimes |w_1\rangle) + (|v\rangle \otimes |w_2\rangle). \tag{4}$$

The tensor product is non-commutative, thus the notation preserves the ordering[7].

The linear operator can be defined over the vector spaces. If we have two linear operators $A$ and $B$, defined on the vector spaces $V$ and $W$, then the linear operator $A \otimes B$ on $V \otimes W$ can be defined as $(A \otimes B)(|v\rangle \otimes |w\rangle) = A|v\rangle \otimes B|w\rangle$, where $|v\rangle \in V$ and $|w\rangle \in W$. In matrix representation, $A \otimes B$ can be expressed as

$$A \otimes B = \begin{bmatrix} A_{11}B & \cdots & A_{1m}B \\ \vdots & \ddots & \vdots \\ A_{m1}B & \cdots & A_{mm}B \end{bmatrix}, \quad (5)$$

where $A$ is an $m \times m$ matrix, and $B$ is an $n \times n$ matrix, hence $A \otimes B$ has dimension $mn \times mn$.

The state $|\psi\rangle$ of an *n*-qubit quantum register is a superposition of the $2^n$ states $|0\rangle, |1\rangle, \ldots, |2^n - 1\rangle$, thus $|\psi\rangle = \sum_{i=0}^{2^n-1} \alpha_i |i\rangle$, with amplitudes $\alpha_i$ constrained to $\sum_{i=0}^{2^n-1} |\alpha_i|^2 = 1$. The state of an *n*-qubit length quantum register is a vector in a $2^n$-dimensional complex vector space, hence if the number of the qubits in the quantum register increases linearly, the dimension of the vector space increases exponentially[7].

A complex vector space $V$ is a Hilbert space $\mathcal{H}$, if there is an *inner product* $\langle \psi | \varphi \rangle$ with $x, y \in \mathbb{C}$ and $|\varphi\rangle, |\psi\rangle, |u\rangle, |v\rangle \in V$ defined by rules of $\langle \psi | \varphi \rangle = \langle \varphi | \psi \rangle^*$, $\langle \varphi | (a|v\rangle + b|v\rangle) \rangle = a \langle \varphi | u \rangle + b \langle \varphi | v \rangle$, and $\langle \varphi | \varphi \rangle > 0$ if $|\varphi\rangle \neq 0$.

If we have vectors $|\varphi\rangle = a|0\rangle + b|1\rangle$ and $|\psi\rangle = c|0\rangle + d|1\rangle$, then the inner product in matrix representation can be expressed as

$$\langle \varphi | \psi \rangle = \begin{bmatrix} a^* & b^* \end{bmatrix} \begin{bmatrix} c \\ d \end{bmatrix} = a^* c + b^* d. \quad (6)$$

The norm of vector $|\varphi\rangle$ can be expressed as $\||\varphi\rangle\| = \sqrt{\langle \varphi | \varphi \rangle}$, and the dual of the vector $|\varphi\rangle$ is denoted by $\langle \varphi |$. The *dual* is a linear operator from the vector space to the complex numbers, defined as[7] $\langle \varphi | (|v\rangle) = \langle \varphi | v \rangle$. The outer product between two vectors $|\varphi\rangle$ and $|\psi\rangle$ can be defined as $|\psi\rangle\langle\varphi|$, satisfying $(|\psi\rangle\langle\varphi|)|v\rangle = |\psi\rangle\langle\varphi|v\rangle$. The matrix of the outer product $|\psi\rangle\langle\varphi|$ is obtained by

usual matrix multiplication of a column matrix by a row matrix, however the matrix multiplication can be replaced by tensor product, since:

$$|\varphi\rangle\langle\psi| = |\varphi\rangle \otimes \langle\psi|. \tag{7}$$

If we have vectors $|\varphi\rangle = a|0\rangle + b|1\rangle$ and $|\psi\rangle = c|0\rangle + d|1\rangle$, the outer product in matrix representation can be expressed as[7]

$$|\varphi\rangle\langle\psi| = \begin{bmatrix} a \\ b \end{bmatrix} \begin{bmatrix} c^* & d^* \end{bmatrix} = \begin{bmatrix} ac^* & ad^* \\ bc^* & bd^* \end{bmatrix}. \tag{8}$$

In Fig. 3. we illustrated the general model of an *n*-length quantum register, where the input state $|\psi_i\rangle$ is either $|0\rangle$ or $|1\rangle$, generally. After the application of a unitary transformation $U$ on the input states, the state of the quantum register can be given by state vector $|\psi\rangle$.

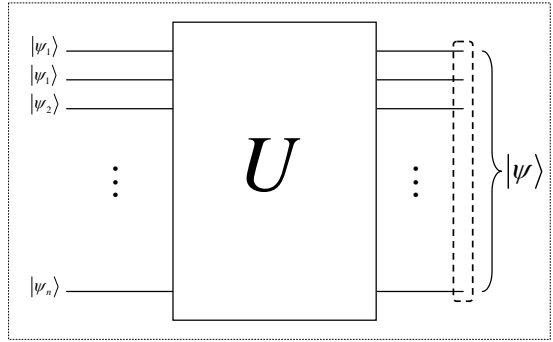

**Fig. 3.** General sketch of an n-length quantum register.

The unitary operator $U$ is a $2^n \times 2^n$ matrix, with – in principle, an infinite number of possible operators. The result of the measurement of state $|\psi\rangle$ results in zeros and ones that form the final result of the quantum computation, based on the *n*-length qubit string stores in the quantum register.

## 2   Motivations behind the Quantum Singular Value Approach

The standard Quantum Fourier Transformation has complexity[7] $\mathcal{O}(N \log N)$. As we show here, the proposed algorithm has the same complexity $\mathcal{O}(N \log N)$, while the precision of this scheme is higher. The main task of our Quantum-SVD algorithm is to find and retain the singular values – retain big coefficients – to eliminate the *input's redundancy*. To compute coefficient vector has complexity $\mathcal{O}(LN)$ approximately, where $L$ denotes the number of the non-zero singular values[10-14]. In general, we use the SVD algorithm where the input has high redundancy, thus the condition $L \ll N$ is often satisfied.

In our Quantum-SVD approach, we will use quantum searching to achieve more precise approximations than using ordinary QFT. The iteration operator of the constructed quantum *oracles*[17] acts for $\mathcal{O}(\sqrt{N})$ times on the initial state[6,7]. Our purpose is to find an *optimal basis,* for which the space generated by basis vectors achieves *better approximations*[7,17].

We present a new method to compute the Quantum Fourier coefficients of *non-uniformly distributed* quantum states. In general, any $m \times n$ matrix $B$, with $m \geq n$ can be factored into three matrices, the $U$ orthogonal $m \times n$ matrix, the $S$ diagonal $n \times n$ matrix, and into an orthogonal $n \times n$ matrix $V$. For complex matrices, $S$ remains real but $U$ and $V$ become unitary. *Unitarity* is the complex analogue of *orthogonality*, and then $B = USV^\dagger$, where $V^\dagger$ is the conjugate transpose of $V$, thus $V^\dagger$ is the complex analogue of $V^T$. The columns of $U$ are the $BB^T$ eigenvectors and the rows of $V$ are the $B^T B$ eigenvectors. The diagonal elements of the $S$ matrix are known as the singular values of $B$, which are the *square roots* of both the non-zero $BB^T$ and $B^T B$. eigenvalues.

In the proposed Quantum-SVD algorithm, we can process *long input sequences at one time*, while in classical systems this must be divided into many small sections on which SVD algorithms act sequentially. The Quantum-SVD algorithm involves expansion of the original data in an *orthogonal* basis

$$x_{ij} = \sum_k b_{ik} v_{jk} = \sum_k u_{ik} s_k v_{jk} = \sum_k c_{ik} e^{\frac{i2\pi jk}{2}}, \qquad (9)$$

thus the normalized vector with unit length is $v_k = e^{\frac{i2\pi jk}{2}}$. The Quantum-SVD algorithm generates the matrix equation $X = USV^T$. Even though $\left\{ v_k = e^{\frac{i2\pi jk}{2}} \right\}$, $k = 0,\ldots,K-1$ is an orthonormal basis, set $\{u_k\}$ is not orthogonal in general[3,2].

The Quantum-SVD algorithm has *conservation property*, thus for any input vector $\vec{x}$ and for output coefficient $\vec{b}$, which is identical to $\sum (x_i)^2 = \sum (|b_i|)^2$. Moreover, our Quantum-SVD algorithm reduces redundancy and the *dimension* of the input data sequence as many of its $\vec{b}$ SVD-

coefficients have values of *close to zero*. The SVD-coefficients denoted by $\vec{b}$, can be discarded without seriously affecting the estimated value of $\|\vec{x}\|^2$, and the main properties of the input data. In the proposed Quantum-SVD algorithm most coefficients are close to zero, and can be discarded. The inverse transformation acts on the "*not close to zero*" coefficients, to restore the original data *approximately*[3].

## 2.1 Basic notations and operators

The matrix *B columns* are $n$ vectors in a complex Hilbert space $\mathcal{H}$, is an outer product between *l*-dimensional subspaces $\mathcal{U} \subseteq \mathcal{H}$ and $\mathcal{V} \subseteq \mathbb{C}^n$. The projector operator for an *l*-dimensional subspace $\mathcal{U} \subseteq \mathcal{H}$ can be expressed as $P_\mathcal{U} = B(B)^*$, and $P_\mathcal{V} = (B)^* B$ for an *l*-dimensional subspace $\mathcal{V} \subseteq \mathbb{C}^n$.

The *outer product* matrix $B$ between *l*-dimensional subspaces $\mathcal{U} \subseteq \mathcal{H}$ and $\mathcal{V} \subseteq \mathbb{C}^n$ can be expressed as

$$B = US_l V^*, \tag{10}$$

where $U$ is a unitary matrix whose first $l$ columns $\{u_i \subseteq \mathcal{H}, 1 \leq i \leq l\}$ are orthonormal basis for $\mathcal{U}$, and $V$ is an $n \times n$ unitary matrix whose first $l$ columns $\{v_i \subseteq \mathcal{H}, 1 \leq i \leq l\}$ are an orthonormal basis for $\mathcal{V}$, while the matrix $S_l$ contains the singular values. Using the definition for $B = US_l V^*$, we can write[1,2,3]

$$B = \sum_{i=1}^{l} u_i v_i^*.$$

The singular value decomposition $B = US_l V^*$ reduces to a sum of $l$ one-dimensional *outer products* $B = \sum_{i=1}^{l} u_i v_i^*$. If we have a vector $u \in \mathcal{U}$, then

$$u = \sum_{i=1}^{l} \langle u_i | u \rangle | u_i \rangle, \text{ and}$$

$$(B)^* u = \sum_{i=1}^{l} \langle v_i | | v_i \rangle \langle u_i | u \rangle | v_i \rangle = \sum_{i=1}^{l} \langle u_i | u \rangle | v_i \rangle. \tag{11}$$

In this case, $(B)^*$ transforms $u$ to a corresponding vector $v \in \mathcal{V}$. If we have $v \in \mathcal{V}$, we can perform the *inverse map*[3] from $\mathcal{V}$ to $\mathcal{U}$ as $Bv = \sum_{i=1}^{l} \langle v_i | v \rangle | u_i \rangle$.

If $u \in \mathcal{H}$, then $(B)^*$ first projects $u$ onto $\mathcal{U}$ and then moves to $\mathcal{V}$, using the outer products. Similarly, for a general $v \in \mathbb{C}^n$, $B$ first projects $v$ onto $\mathcal{V}$ and then takes it to $\mathcal{U}$. The $l$-dimensional outer product matrix $B$ is a *partial distance-preserving transformation*[3] between the subspaces $\mathcal{U} \subseteq \mathcal{H}$, and $\mathcal{V} \subseteq \mathbb{C}^n$. If we assume, that $v, v' \in \mathcal{V}$, $u = Bv$ and $u' = Bv'$, then

$$\begin{aligned}\langle u|u'\rangle &= u^*u' = \langle u||u'\rangle = u^*B^*Bu' = \langle v|\langle B||B\rangle|v'\rangle \\ &= v^*P_\mathcal{V}v' = \langle v|P_\mathcal{V}|v'\rangle = v^*v' = \langle v||v'\rangle = \langle v|v'\rangle,\end{aligned} \quad (12)$$

where $P_\mathcal{V}$ is an $l$-dimensional orthogonal projector onto an $l$-dimensional subspace $\mathcal{V} \subseteq \mathbb{C}^n$ with an orthonormal basis $\{v_i \in \mathbb{C}^n, 1 \leq i \leq l\}$, consisting of the first $l$ columns of $V$.

As we can conclude, inner products and a fortiori squared norms and distances[3] can be preserved. If $u, u' \in \mathcal{U}$, then $\langle (B)^*u|(B)^*u'\rangle = \langle u|u'\rangle$. Moreover, if $u, u' \notin \mathcal{U}$ or $v, v' \notin \mathcal{V}$, then the inner products *are not preserved*[3].

## 2.2 Error probability of the QFT based interpolation

The decomposition of *singular values* is much more complex than a Hadamard transform[7], and it lets us perform more effective extraction of data from the input than a simple Hadamard rotation. The *Quantum Fourier Transform* is a unitary operation on $n$ qubits can be given by[7]

$$QFT|x\rangle : F|x\rangle = \frac{1}{\sqrt{N}} \sum_{y=0}^{N-1} e^{\frac{i2\pi xy}{N}} |y\rangle = \frac{1}{\sqrt{2^n}} \sum_{y=0}^{2^n-1} e^{\frac{i2\pi xy}{2^n}} |y\rangle, \quad (13)$$

where $xy$ is a "decimal" multiplication of numbers $x$ and $y$, which numbers are represented by quantum registers $|x\rangle = |x_{n-1}\rangle \otimes |x_{n-2}\rangle \otimes |x_{n-3}\rangle \otimes ... \otimes |x_0\rangle$ and $|y\rangle = |y_{n-1}\rangle \otimes |y_{n-2}\rangle \otimes |y_{n-3}\rangle \otimes ... \otimes |y_0\rangle$. The effect of transformation $F$ on the basis vectors could be extended to other vectors[7]. If we know what transformation $F$ does to the basis vectors $|x\rangle$, then we can figure out what $F$ does to any other vector

$$\sum_{i=0}^{N} f(x)|x\rangle, \quad (14)$$

where $f(x) = \dfrac{1}{N} = \dfrac{1}{2^n}$. The effect of transformation $F$ on the basis vectors can be extended to other vectors. Let assume, we have an input vector $\sum_{x=0}^{N-1} f(x)|x\rangle$, and we apply the singular value decomposition transformation $F$ to it, thus

$$F\left(\sum_{x=0}^{N-1} f(x)|x\rangle\right) = \sum_{x=0}^{N-1} f(x) F|x\rangle = \dfrac{1}{\sqrt{N}} \sum_{y=0}^{N-1} \sum_{x=0}^{N-1} f(x) e^{\frac{i 2\pi x y}{N}} |y\rangle. \tag{15}$$

From this decomposition formula the $y^{th}$ component of $QFT|x\rangle$ is

$$F_y(f) = \dfrac{1}{\sqrt{N}} \sum_{x=0}^{N-1} f(x) e^{\frac{i 2\pi x y}{N}}. \tag{16}$$

In the quantum decomposition transform for the $i^{th}$ input state $|x_i\rangle$ we want to calculate $F(x_i) = \dfrac{1}{\sqrt{N}} \sum_{y=0}^{N-1} f(x_i) e^{\frac{i 2\pi x_i y}{N}} |y\rangle$, where $e^{\frac{i 2\pi x_i y}{N}} = e^{\frac{i 2\pi x_i y}{2^n}}$ is a periodic function in $xy$, with period $2^n = N$.

We can define $F^{-1}$ as the inverse transformation[7] of QFT, where the only difference is in the change of the sign in front of the imaginary unit $i$:

$$F^{-1}|x\rangle = \dfrac{1}{\sqrt{2^n}} \sum_{y=0}^{2^n-1} e^{-\frac{i 2\pi x y}{2^n}} |y\rangle. \tag{17}$$

Let's introduce the notation $\theta = \dfrac{y}{2^n}$ for input state. The probability that measurement results in outcome $|y'\rangle$ is therefore

$$p_{y'} = \left| \dfrac{1}{2^n} \sum_{x=0}^{2^n-1} f(x) e^{\frac{i 2\pi x (y-y')}{2^n}} \right|^2 = \left| \dfrac{1}{2^n} \sum_{x=0}^{2^n-1} e^{i 2\pi x (\theta - y'/2^n)} \right|^2, \tag{18}$$

where $f(x) = \dfrac{1}{N} = \dfrac{1}{2^n}$ for each $y' \in \{0, \ldots, 2^n - 1\}$. However, in general we cannot assume that $\theta$ has this special form, thus let assume that $\theta \neq \dfrac{y}{2^n}$, so $e^{i2\pi(\theta - y'/2^n)} \neq 1$, for every integer $y'$. In the case $\theta = \dfrac{y}{2^n}$, once we know $y' = y$, we know $\theta$ as well, because in this special case $\theta = \dfrac{y}{2^n}$.

The *inverse* Quantum Fourier transform[7] gives us the inner product between the input state $|\psi_y\rangle = \dfrac{1}{\sqrt{2^n}} \sum_{x=0}^{2^n-1} e^{i2\pi \frac{xy}{2^n}} |x\rangle$ and the *orthogonal bases* $|\psi_{y'}\rangle = \dfrac{1}{\sqrt{2^n}} \sum_{x=0}^{2^n-1} e^{i2\pi \frac{xy'}{2^n}} |x\rangle$ for each $y' \in \{0,...,2^n-1\}$, thus

$$\langle \psi_y | \psi_{y'} \rangle = \dfrac{1}{2^n} \sum_{x=0}^{2^n-1} e^{i2\pi \frac{x(y-y')}{2^n}} |x\rangle. \tag{19}$$

Using the *geometric series* formula $\sum_{k=0}^{n-1} x^k = \dfrac{x^n - 1}{x - 1}$ for $x \neq 1$, and $\sum_{k=0}^{n-1} 1^k = n$ along with that $e^{i2\pi l} = 1$ for any integer $l$, we obtain $\langle \psi_j | \psi_{j'} \rangle = 1$ if $y = y'$, and 0 otherwise. In this case the set of input states $\{|\psi_0\rangle, |\psi_1\rangle, ..., |\psi_{2^m-1}\rangle\}$ is indeed orthonormal[7]. However, if we use the formula for the geometric series $\sum_{k=0}^{n-1} x^k = \dfrac{x^n - 1}{x - 1}$ again in the case of $e^{i2\pi(\theta - y'/2^n)} \neq 1$, for every integer $y'$, and for $x \neq 1$, then

$$p_{y'} = \dfrac{1}{2^{2n}} \left| \dfrac{e^{i2\pi(2^n \theta - y')} - 1}{e^{i2\pi(\theta - y'/2^n)} - 1} \right|^2. \tag{20}$$

The probability $p_{y'}$ is large, for values of $y'$ that satisfy $y'/2^n \approx \theta$, and small otherwise[1,4].

The error produced by the proposed Quantum-SVD method is some orders lower than error produced with the interpolation approach. Our Quantum-SVD algorithm is based on singular value decomposition, and in comparison with the Quantum Fourier interpolation approach, it has a lower computational complexity. The Quantum-SVD approach provides advantages in terms of computational structure, based on QFT and multiplications.

## 3 Quantum-SVD based Approximation

Let assume we have $\{\psi_i, 1 \leq i \leq n\}$ a set of $n$ vectors in an $l$-dimensional subspace $\mathcal{U}$ of a Hilbert space $\mathcal{H}$. The vectors $\psi_i$ form a *tight approximation set*[4,6-7] for $\mathcal{U}$, if there exists a constant $\beta > 0$ such that $\sum_{i=1}^{n} |\langle x | \psi_i \rangle|^2 = \beta^2 \|x\|^2$, for all $x \in \mathcal{U}$.

If $\beta = 1$ the *approximation vector*[6-7] is normalized, otherwise it is $\beta$-scaled. The vectors $|\psi_i\rangle$ form an *approximation set*[3,4] for $\mathcal{U}$, if there exist constants $\alpha > 0$ and $\beta < \infty$ such that $\alpha^2 \|x\|^2 \leq \sum_{i=1}^{n} |\langle x | \psi_i \rangle|^2 \leq \beta^2 \|x\|^2$, for all $x \in \mathcal{U}$. The lower bound $\alpha^2 \|x\|^2$ ensures that the vectors $|\psi_i\rangle$ span $\mathcal{U}$, thus $n \geq l$. If $n < \infty$, then the right hand inequality is always satisfied with $\beta^2 = \sum_{i=1}^{n} \langle \psi_i | \psi_i \rangle$, thus any finite set of vectors that spans $\mathcal{U}$ is an *approximation set*[4,6-7] for $\mathcal{U}$, and any basis for $\mathcal{U}$ is an approximation for $\mathcal{U}$.

In our method we use the fact that any *orthonormal* basis for $\mathcal{U}$ is a normalized approximation vector set for $\mathcal{U}$, however there also exists tight approximation vectors for $\mathcal{U}$ with $l < n$, which are necessarily *not linearly independent*[4].

The *redundancy* of the tight approximation vector set is defined as $\rho = n/l$, since[4]

$$\sum_{i=1}^{n} |\langle x | \psi_i \rangle|^2 = \beta^2 \|x\|^2 = \sum_{i=1}^{n} x^* \psi_i \psi_i^* x = \sum_{i=1}^{n} \langle x \| \psi_i \rangle \langle \psi_i \| x \rangle = \langle x | \left( \sum_{i=1}^{n} |\psi_i\rangle \langle \psi_i| \right) | x \rangle.$$

Using that $\sum_{i=1}^{n} |\langle x | \psi_i \rangle|^2 = \beta^2 \|x\|^2$ holds for all $x \in \mathcal{U}$, we can write that $\sum_{i=1}^{n} |\psi_i\rangle\langle \psi_i| = \beta^2 P_\mathcal{U}$, where $P_\mathcal{U} = MM^*$, and matrix $M$ is a measurement matrix[3] with $n$ columns in $\mathcal{H}$ for states in the subspace $\mathcal{U} \subseteq \mathcal{H}$. When the vectors $\psi_i \in \mathcal{U}$ satisfy $\sum_{i=1}^{n} |\psi_i\rangle\langle \psi_i| = \beta^2 P_\mathcal{U}$, then $\langle x | \left( \sum_{i=1}^{n} |\psi_i\rangle\langle \psi_i| \right) | x \rangle$ implies that $\sum_{i=1}^{n} |\langle x | \psi_i \rangle|^2 = \beta^2 \|x\|^2$ is satisfied for all $x \in \mathcal{U}$. Thus a set of $n$ vectors $\psi_i \in \mathcal{U}$

forms a tight approximation set for $\mathcal{U}$ if and only if the vectors satisfy[1,4] equation $\sum_{i=1}^{n}|\psi_i\rangle\langle\psi_i| = \beta^2 P_{\mathcal{U}}$ for some $\beta > 0$.

Comparing $\sum_{i=1}^{n}|\psi_i\rangle\langle\psi_i| = \beta^2 P_{\mathcal{U}}$ and $MM^* = P_{\mathcal{U}}$ we conclude that a set of vectors $\psi_i \in \mathcal{U}$ forms a $\beta$-scaled tight approximation set[6-7] for $\mathcal{U}$ if the scaled vectors $\beta^{-1}\psi_i$ are the measurement vectors of a rank-one positive operator-valued (POVM) measurement[3,7] on $\mathcal{U}$. The vectors $\psi_i \in \mathcal{U}$ form a normalized tight approximation[3,6-7] for $\mathcal{U}$ if they are measurement vectors of a rank-one POVM on $\mathcal{U}$.

Using this relationship between rank-one quantum measurements and tight approximation set, we can define approximation matrices in analogy to the measurement matrices[3,7] of quantum mechanics.

### 3.1 Mathematical background

Using the singular value decomposition of matrix $B$, we can write that $B = US_l V^*$, where $S_l$ is the *l*-dimensional matrix of *singular values*. If $k \geq n$, then $B$ has at least as many rows as columns, and we construct orthogonal matrix $\tilde{B}$ by extending the *I* identity matrix along the diagonal, thus $\tilde{B} = US_l V^*$. The *left* and *right* unitary matrices in the SVD of matrix $B$ and orthogonal matrix $\tilde{B}$ are the same, and are equal to $U$ and $V$. If $k=n$, then $S_n = I_n$ and $\tilde{B} = UI_n V^* = UV^*$. Finally, if $k<n$, we first replace the *left* unitary matrix $U$ by $\tilde{U}$, and thus replace $k$ by $\tilde{k} = n$, then $\tilde{U}$ is an $n \times n$ unitary matrix whose first $l$ columns are the $\mathcal{U}$ basis, where we append *n-k* zeros[3,4,5] to each basis vector $u_i$.

In the QFT algorithm each *non-uniform* angle $\theta \neq \dfrac{y}{2^n}$ is obtained by *interpolating* the orthogonal $\theta = \dfrac{y}{2^n}$ coefficients in the neighborhood of the considered *non-uniform angle*[7]. In our Quantum-SVD based approach, each *non-uniform* angle $\theta \neq \dfrac{y}{2^n}$ in the interval between two uniform angles $\theta = \dfrac{y}{2^n}$ is approximated by a *small set of orthogonal vectors*. We use the fact, that all non-uniform angles can be obtained by modulating a *non-uniform angle* of a single interval[1,2,3].

The main steps of our Quantum-SVD based algorithm are:
1. The input quantum register is scaled by a quantum oracle[17] according to a small set of vectors[1-4,10-13].
2. The quantum register is transformed by Quantum Fourier Transformation[15-17,22-26].
3. By defining a quantum oracle[7,17], the Quantum Fourier coefficients are linearly combined, in analogy with the interpolation techniques.

## 3.2 Construction of quantum approximation matrices

Using SVD, the approximation matrix $B$ of a $\beta$-scaled *tight* approximation for an $l$-dimensional subspace $\mathcal{U} \subseteq \mathcal{H}$ may be expressed as $B = \beta\left(US_lV^*\right)$, where $U$ is a unitary matrix whose first $l$ columns $\{u_i, 1 \leq i \leq l\}$ are an orthonormal basis for $\mathcal{U}$, while $\mathcal{V}$ is an $n \times n$ unitary matrix whose first $l$ columns $\{v_i, 1 \leq i \leq l\}$ are an orthonormal basis for $\mathcal{V}$, and $S_l$ is the singular value matrix. The matrix $B$ can be expressed as $B = \beta \sum_{i=1}^{l} u_i v_i^*$. The approximation matrix $B$ of a $\beta$-scaled *tight* approximation is an *outer product,* if restricted to $\mathcal{V}$ and scaled by $\beta^{-1}$, and the columns of matrix $B$ are $n$ vectors in $H$, represents an orthogonal basis for $U$ iff its rank[4] is $n$.

$B$ is an *orthogonal approximation matrix* if its rank is $n$, thus $B = \beta\left(US_lV^*\right)$ and $\left(B\right)^* B = \beta^2 I_n$, and all vectors have *squared norm* $\beta^2$. Suppose we have a given approximation matrix $B$ in $\mathcal{U}$. We can give a concrete construction of an *orthogonal* approximation matrix $\tilde{B}$ in $\tilde{\mathcal{U}} \supseteq \mathcal{U}$ such that $P_{\mathcal{U}}\tilde{B} = B$.

Consider we have approximation vector set $\{|\psi_1\rangle = \alpha_1|0\rangle + \beta_1|1\rangle, |\psi_2\rangle = \alpha_2|0\rangle + \beta_2|1\rangle, ..., |\psi_N\rangle = \alpha_N|0\rangle + \beta_N|1\rangle\}$, then the approximation matrix is

$$B = \begin{pmatrix} \alpha_1 & \alpha_2 & ... & \alpha_N \\ \beta_1 & \beta_2 & ... & \beta_N \end{pmatrix}. \tag{21}$$

The matrix $B$ is an approximation matrix of a *tight approximation set*, since $B(B)^* = I_2$. To construct an *orthogonal* approximation *matrix* $\tilde{B}$ such that $P_{\mathcal{U}}\tilde{B} = B$, we use the singular value decomposition $B = USV^*$. We express $B$ as $B = USV^*$, and let $u_i$ and $v_i$ denote the columns of $U$ and $V$ respectively[4].

Assume that $\mathcal{H}$ is finite dimensional, and $k = \dim \mathcal{H}$. If $k \geq n$, then $B$ has at least as many rows as columns, while if $k<n$ matrix $B$ has more columns than rows.

We define the orthogonal matrix $\tilde{B}$ as $\tilde{B} = \sum_{i=1}^{n} u_i v_i^*$, then $\tilde{\mathcal{U}} \subset \mathcal{H}$ is the $n$-dimensional subspace spanned by $\{u_i, 1 \leq i \leq n\}$. The projection of $\tilde{B}$ onto $\mathcal{U}$ is[3]

$$P_{\mathcal{U}} \tilde{B} = \sum_{j=1}^{m} u_j v_j^* \sum_{i=1}^{n} u_i v_i^* = \sum_{i=1}^{m} u_i v_i^* = B. \tag{22}$$

The columns of $\tilde{B}$ are orthonormal, since $\left(\tilde{B}\right)^* \tilde{B} = \sum_{j=1}^{m} v_j u_j^* \sum_{i=1}^{n} u_i v_i^* = \sum_{i=1}^{n} v_i v_i^* = I_n$. If $k<n$, we insert $\mathcal{U}$ in an $n$-dimensional space $\tilde{\mathcal{U}}$ in an expanded complex Hilbert space $\tilde{\mathcal{H}} \supset \mathcal{H}$, and let $\{u_i, 1 \leq i \leq n\}$. be an orthonormal basis for $\tilde{\mathcal{U}}$ of which the first $m$ vectors are the $\mathcal{U}$-basis[5]. Then we can use $\tilde{u}_i$ in place of $u_i$.

## 4  Description of Quantum-SVD Approximation

The proposed Quantum-SVD algorithm consists of *four* main steps[1-4,10-13,16,17]:

(1) *Computing the interpolation coefficients* using Quantum Fourier Transformation as $P_l = \left[\left(\sigma_{1...0}\right)^i QFT^{-1} \tilde{\mathbf{E}}_l\right]^T U_L$ for *l=2i*, and $P_l = \left[\Upsilon_{0...1}\left(\sigma_{1...0}\right)^i QFT^{-1} \tilde{\mathbf{E}}_l\right]^T U_L$ for *l=2i-1*, using quantum oracle[17] $\mathbf{O}_{Precomp.}$, where $P_l$ is a real $\left[K_l \times M\right]$ matrix, which combines the column vectors of $U_L$ to get the best approximation[10-13] of $\tilde{\mathbf{E}}_{l,0}$ (Following the notations of Refs. [10], [12], [13]). Matrix $\tilde{\mathbf{E}}_{l,0}$ refers to randomly distributed angles in $\left[0, 1/2\right)$. Operator $\sigma_{1...0}$ defines circular shift unitary operation, and $\Upsilon_{0...1}$ is a complex conjugation only if the result of QFT transform $QFT^{-1} \tilde{\mathbf{E}}_l$ is a real vector. Matrix $U_L$ denotes the first $L$ columns of matrix $U$, and $QFT^{-1} \tilde{\mathbf{E}}_0 = U_L S_L V_L$. The operator $\Upsilon$ acts as a *complex conjugation* only if it is applied to the *Quantum Fourier* transform of a real vector[1-4,10-13,16], thus $QFT^{-1}\mathcal{X}$ is real if $\mathbb{R}[\mathcal{X}(0,\cdot)] = 0$. We also introduce the operator $\tilde{\mathbf{E}} = \mathcal{X} - i|\sigma_0\rangle\langle\sigma_0|\mathbb{R}[\mathcal{X}]$. Following [10-14], we define the row vector $\mathbf{D} = i\mathbb{R}[\mathcal{X}(0,\cdot)]$ of size $K$, the column vector $|X\rangle$ of size $K$ representing the non-unitary QFT transformation of the state of the input quantum register $x$ results[1,2] $X = \mathcal{X}^\dagger x = \tilde{\mathbf{E}}^\dagger x + \mathbf{D}^\dagger x_0$.

(2) *Computing* $L$ *scaling*[1-4,10-13] $\Delta = diag(x)W = diag(x)QFT^{-1}U_L$, *using quantum oracle*[17] $\mathbf{O}_{Scale}$, where $diag(x)$ is a matrix having $x$ as main diagonal, and $QFT^{-1}U_L$ is the inverse QFT transformed vectors[10-13] of $U_L$.

(3) *Calculating* $N \times L$ *scalar products* $\mathbf{Q} = QFT\Delta$, *using Quantum Fourier Transformation and quantum oracle*[17] $\mathbf{O}_{Outer\ prod.}$, where $\Delta = diag(x)QFT^{-1}U_L$.

(4) *Computing the output* $X = \mathcal{X}^\dagger x = \tilde{\mathbf{E}}^\dagger x + \mathbf{D}^\dagger x_0$ *of the input quantum register $x$, by interpolating the scalar products*[1-4,10-13] $\mathbf{Q} = QFT\Delta$, *by* $\tilde{X}_{(l)} \simeq P_l Q_i^\dagger$ *for l=2i, and* $\tilde{X}_{(l)} \simeq P_l Q_{-i}^T$ *for l=2i-1 using quantum oracle* $\mathbf{O}_{Interpol.}$, *where* $Q_i$ *is the i-th row of* $\mathbf{Q} = QFT\Delta$, *and* $P_l = \left[\left(\sigma_{1\ldots 0}\right)^i QFT^{-1}\tilde{\mathbf{E}}_l\right]^T U_L$ *for l=2i, and* $P_l = \left[\Upsilon_{0\ldots 1}\left(\sigma_{1\ldots 0}\right)^i QFT^{-1}\tilde{\mathbf{E}}_l\right]^T U_L$ *for l=2i-1*.

In Fig. 4. we illustrate the main steps of the Quantum-SVD approach:

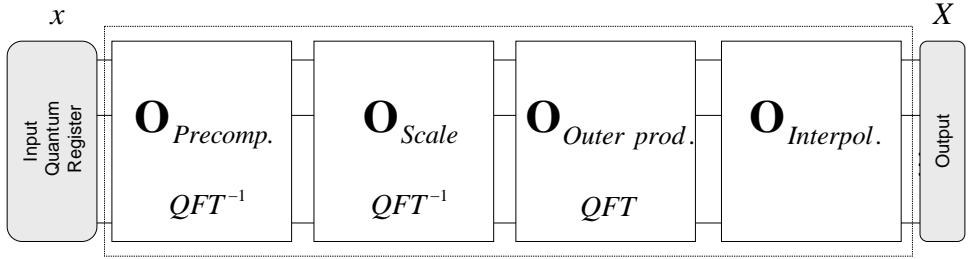

**Fig. 4.** Decomposition of the Quantum-SVD algorithm.

To apply the Quantum-SVD algorithm we use the *quantum search algorithm*[7,17] to obtain the results of the multiplications, decompositions, inner and outer products, and the QFT transformation.

### 4.1 Implementation of quantum searching

The general unitary iteration $G$ of quantum search algorithm[7,17] can be given by

$$G = \left(2|\psi\rangle\langle\psi| - I\right)\left(U_L\right)^\dagger \left(O_C\right)^\dagger O_F O_C U_L, \qquad (23)$$

where $U_L$ is the unitary quantum loading operation, $O_C$ is the matching oracle, while $O_F$ is the judge function oracle[7,17].

The output of $O_F$ is 1 if output of $O_C$ satisfies the given condition, otherwise $O_F = 0$. The quantum state $|\psi\rangle$ contains all the index states $|\psi\rangle = \frac{1}{\sqrt{N}} \sum_{i=0}^{N-1} |i\rangle$, where $|i\rangle$ is the $i$-th index state in the quantum register[7]. The iteration operator $G$ acts for $\mathcal{O}\left(\sqrt{N}\right)$ times on initial state $\frac{1}{\sqrt{N}} \sum_{i=0}^{N-1} |i\rangle|0\rangle$, where the state $|i\rangle$ is in quantum register $A$, while the initial state $|0\rangle$ is in quantum register $B$. Thus, we have to apply the iteration operator $G$ for $\mathcal{O}\left(\sqrt{N}\right)$ times to find the optimal solution[7].

The quantum search algorithm has complexity $\mathcal{O}\left(\sqrt{N}\right)$, and we can use it to search for an element with a unique index $i_0$ in a list of $N$ unsorted elements[7]. In quantum searching, the records use a superposition index state $\frac{1}{\sqrt{N}} \sum_{i=0}^{N-1} |i\rangle$ to find index $i_0$ of database record $r_{i_0}$, where $r_{i_0}$ is the answer that we have searched for[7].

### 4.2 Computing basic Fourier coefficients

We consider an $N \times K$ matrix $\mathcal{X}$, whose columns are *non-uniform* QFT exponentials, where $K$ is the cardinality of the set of *non-uniform* angles[1-4,10-13]. Thus, the elements of matrix $\mathcal{X}(n,k)$ are

$$\mathcal{X}(n,k) = e^{2\pi i\left(\frac{y_k}{2^n} + \varepsilon\right)} = e^{2\pi i(n-N/2)\frac{\phi_k^{(l)}}{N}} = e^{2\pi i \theta}, \qquad (24)$$

where $n = 0...N-1$, and $k = 0...K-1$. The non-uniform angle set $\theta_k$ is partitioned in $2N$ subsets corresponding to contiguous angle intervals[10-13]

$$\phi_k^{(l)} \in \left[\frac{l}{2}, \frac{l}{2} + \frac{1}{2}\right), \qquad (25)$$

where $l = 0,\ldots,2N-1$ and whose cardinalities are $K_l$. The set of non-uniform QFT exponentials corresponding to each interval is considered as

$$\mathcal{X}_l(n,k) = e^{2\pi i \left(\frac{y_k}{2^n}+\varepsilon\right)^{(l)}} = e^{2\pi i (n-N/2)\frac{\phi_k^{(l)}}{N}} = e^{2\pi i \theta_k^{(l)}}. \qquad (26)$$

Following [10-14], the matrix $\mathcal{X}_l$ for *even l* could be obtained by modulating the column vectors of a non-uniform *Quantum-Fourier* matrix referred to a suitable set of angles $\phi_k \in [0, 1/2)$, which will be represented as $\mathcal{X}_{l,0}$. For *odd l*, $\mathcal{X}_l$ could be obtained from a non-uniform QFT matrix, whose angles refer to $\phi_k \in [-1/2, 0)$, which can also be obtained by conjugating[10-13] a suitable matrix $\mathcal{X}_{l,0}$.

The modulation can be converted in circular shift by passing in the Quantum Fourier domain[4,10-14] by $\tilde{\mathbf{E}}_l = (-1)^i QFT(\sigma_{1\ldots 0})^{-i} QFT^{-1}\tilde{\mathbf{E}}_{l,0}$, for *l=2i*, and $\tilde{\mathbf{E}}_l = (-1)^i QFT(\sigma_{1\ldots 0})^{-i} QFT^{-1}\tilde{\mathbf{E}}_{l,0}^* \sigma_{N-1\ldots 0}$, for *l=2i-1*. Each matrix $\tilde{\mathbf{E}}_{l,0}$ refers to randomly distributed angles in range $[0, 1/2)$, thus in order to find a single basis for the column vectors of all matrices $\tilde{\mathbf{E}}_{l,0}$, we consider a generic matrix[1-4,10-13] $\mathcal{X}_0$.

The *singular values* decay exponentially, thus we could consider a reduced number $L$ of singular values[1-4,10-13] and neglect the others by:

$$B = QFT^{-1}\tilde{\mathbf{E}}_0 \simeq U_L S_L V_L^T, \qquad (27)$$

where $U_L$ and $V_L$ are constituted by the first $L$ columns of $U$ and $V$, and $S_L$ is a square matrix constituted by the first $L$ rows and columns of $S$.

Fig. 5. illustrates the entries of transformation $B = QFT^{-1}\tilde{\mathbf{E}}_0$, the column vectors are correlated, and the entries have small rank.

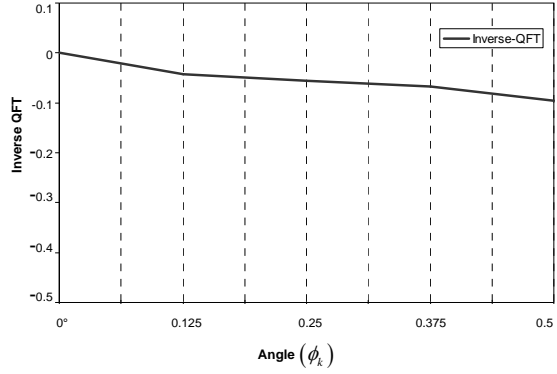

**Fig. 5.** Inverse Quantum Fourier Transformation of non-uniform QFT exponentials.

After the inverse QFT transformation it can be concluded, that the entries $B = QFT^{-1}\tilde{\mathbf{E}}_0$ are very correlated, and the rank becomes small.

### 4.3 Oracle-based Approximation of Quantum Fourier coefficients

We construct oracle $\mathbf{O}_{Precomp.}$ to compute matrix $P_l$, using the orthogonal columns of matrix $U_L$ we get $P_l = \left[\left(\sigma_{1...0}\right)^i QFT^{-1}\tilde{\mathbf{E}}_l\right]^T U_L$, for $l = 2i$, and $P_l = \left[\Upsilon_{0...1}\left(\sigma_{1...0}\right)^i QFT^{-1}\tilde{\mathbf{E}}_l\right]^T U_L$ for $l = 2i - 1$.

The column vectors of each $\tilde{\mathbf{E}}_l$ are approximated by linearly combining the columns of $QFT\left(\sigma_{1...0}\right)^{-i} U_L$, or $QFT\left(\sigma_{1...0}\right)^{-i} \Upsilon_{0...1} U_L$ through the rows of matrix $P_l$. The oracle $\mathbf{O}_{Precomp.}$ is a *simple parallel circuit*, and the running time of the oracle $\mathbf{O}_{Precomp.}$ can be regarded as few times of multiplications, which is denoted by $t_{mult}$. The addition is more fast than the multiplication, thus the computational complexity of oracle $\mathbf{O}_{Precomp.}$ is $\mathcal{O}(1)t_{mult}$, where the computational complexity of $\mathbf{O}_{Precomp.}$ could be given by the computational cost $\mathcal{O}(N \log N)$ of calculating the interpolation coefficients, which consists of *inverse Quantum Fourier transforms and interpolation*[7].

**4.4 Oracle for computing scaling**

In order to compute the output $X = \mathcal{X}^\dagger x = \tilde{\mathbf{E}}^\dagger x + \mathbf{D}^\dagger x_0$, where matrix $\mathcal{X}$ contains the *non-uniform* QFT exponentials. We construct *oracle*[7,17] $\mathbf{O}_{Scale}$ to calculate the scalar product between $x$ and the columns of $QFT(\sigma_{1\ldots 0})^{-i} U_L$ and $QFT(\sigma_{1\ldots 0})^{-i} (\Upsilon_{0\ldots 1}) U_L$.

Following the notations of [10-14], firstly, $x$ must be scaled by the *inverse Quantum Fourier* transformed vectors of $U_L$, so we construct $\mathbf{W} = QFT^{-1} U_L$, and $\Delta = diag(x)\mathbf{W} = diag(x) QFT^{-1} U_L$, where $diag(x)$ is a matrix with $x$ as main diagonal[3]. The oracle $\mathbf{O}_{Scale}$ can be constructed by a simple quantum circuit, and the running time of the oracle $\mathbf{O}_{Scale}$ can be regarded as few times of multiplications[7,17]. The computational complexity of oracle $\mathbf{O}_{Scale}$ is $\mathcal{O}(1) t_{mult}$., where $t_{mult}$ denotes times of multiplications, since the $\Delta = diag(x) QFT^{-1} U_L$ can be computed in a very efficient way. The computational complexity of $\mathbf{O}_{Scale}$ could be given by the computational cost of calculating the scaling[1-4,10-13] (using the quantum oracle $\mathbf{O}_{Scale}$) and the cost $\mathcal{O}(N \log N)$ of *inverse Quantum Fourier* transforms[1,2,4].

**4.5 Oracle for computing outer products and interpolation**

We construct oracle $\mathbf{O}_{Outer\ prod.}$ to compute the *QFT transforms* of the column vectors of $\Delta$ produce. To compute matrix $\mathbf{Q}$, we design a simple quantum circuit to compute the $N \times L$ scalar products $\mathbf{Q} = QFT\Delta$, where each row of $\mathbf{Q}$ will be headed by $\mathbf{Q}_i$. The computational complexity of $\mathbf{O}_{Outer\ prod.}$ could be given by the computational cost $\mathcal{O}(N \log N)$ of the standard QFT complexity[3,4,10-14].

Finally, we construct oracle $\mathbf{O}_{Interpol.}$ to compute $\tilde{X}_{(l)} = \tilde{\mathbf{E}}_l^\dagger x$. The oracle $\mathbf{O}_{Interpol.}$ first computes the *inner* product between the $K_l \times L$ matrix and a column vector of size $L$, thus $\tilde{X}_{(l)} \simeq P_l \mathbf{Q}_i^\dagger$ for $l = 2i$, and $\tilde{X}_{(l)} \simeq P_l \mathbf{Q}_{-i}^T$ for $l = 2i - 1$. The oracle $\mathbf{O}_{Interpol.}$ is a parallel quantum circuit, and the time complexity of the *inner* products can be given by the cost of *interpolation*[1-4,10-13].

The computational complexity of $\mathbf{O}_{Interpol.}$ could be given by the computational cost $\mathcal{O}(N \log N)$ of calculating *Quantum Fourier* transforms[6].

## 5  Results

### 5.1  Computational complexity

According to the decomposition of the algorithm (see Fig. 4), the Quantum-SVD algorithm has computational complexity

$$\sqrt{2LN} + \left(4LN \log N\right) + \sqrt{2LN\left(\frac{K}{N}\right)} \approx \mathcal{O}(N \log N), \tag{28}$$

where $K$ is the cardinality[1-4,10-13] of the set of *non-uniform angles*[1-4,10-13].

- The first term $\sqrt{2LN}$ derives from $\Delta = diag(x)W = diag(x)QFT^{-1}U_L$, where $L$ products between a real and complex vector of size $N$ are computed, it implies the $\sqrt{2LN}$ time[1-4,10-13].
- The term $4LN \log N$ derives from the Quantum Fourier Transforms $Q = QFT\Delta$.
- The term $\sqrt{2LN\left(\frac{K}{N}\right)}$ derives from the computation of $\tilde{X}_{(l)} \simeq P_l Q_i^\dagger$ for *l=2i*, and $\tilde{X}_{(l)} \simeq P_l Q_{-i}^T$ for *l=2i-1*.

Using analogous considerations for *QFT based* interpolating techniques[10-13] it results in

$$\sqrt{N} + \left(8N \log N\right) + \sqrt{4LN\left(\frac{K}{N}\right)}\sqrt{\lambda} \approx \mathcal{O}(N \log N), \tag{29}$$

where the factor $\lambda > 1$ is the *effectiveness parameter*[10-13], which takes into account that, the same value of *L*, the SVD-based approach performs more precise. From this result we get a gain factor of $\sqrt{\lambda} \approx \sqrt{1.4}$, thus our increment in the *precision* is $\lambda \approx 1.4 \approx 2^{2\delta}$, and the increment factor[10-13] is

$$\delta = \log_2 \sqrt{\lambda} = \log_2 \sqrt{1.4} \approx 0.24. \tag{30}$$

In Fig. 6 we compared the approximation errors of non-uniform quantum states given by ordinary QFT and the proposed Quantum-SVD approach.

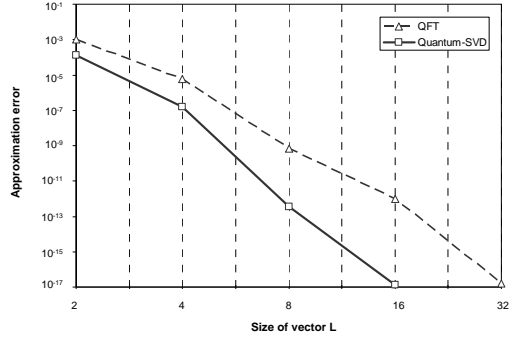

**Fig. 6.** Error of classical QFT algorithm and Quantum-SVD approach as the size of vector *L*.

It can be concluded that the estimated error of the proposed QFT approach becomes some orders lower as the size of *L* increases.

### 5.2 Precision of Quantum-SVD based approximation

In this section we demonstrate that the Quantum-SVD algorithm has better precision in comparison to the standard QFT. Using the proposed scheme an improved approximation of input angle $\theta$ can be achieved. In QFT the *k* bits of precision can be achieved if[7] $n = k + 2$.

For a general input state set, we cannot assume that $\theta$ has the special form $\theta = \dfrac{y}{2^n}$, and if $\theta \neq \dfrac{y}{2^n}$, then $e^{i2\pi(\theta - y'/2^n)} \neq 1$, for every integer $y'$. We have to calculate with the *error probability* $\varepsilon$ and the best possible output $y'$. In our Quantum-SVD algorithm, we can achieve *more precision*, using the same number of quantum states *n*. Using *n* quantum states in Quantum-SVD is equivalent to applying $n + \delta$ states in the QFT algorithm to approximate $\theta$. In QFT minimum outcome, which is accurate to *k+2* bits of precision, occurs with probability $4/\pi^2$.

Using our Quantum-SVD algorithm's *increment factor* $\delta$ over ordinary QFT, the probability of obtaining the best possible $y'$ to approximate $\theta \neq \dfrac{y}{2^{n+\delta}}$ is $e^{i2\pi\theta} = e^{i2\pi(y'/2^{n+\delta} + \varepsilon)}$, for some real number $\varepsilon$ with $|\varepsilon| \leq 2^{-(n+\delta+1)}$. Using this error threshold, we can approximate $\theta \neq \dfrac{y}{2^{n+\delta}}$ in the form[3,6,8] $\theta = \dfrac{y}{2^{n+\delta}} + \varepsilon$.

Assuming that $y'$ satisfies this equation, a *lower bound*[8] on probability $p_{y'} = \frac{1}{2^{2n}} \left| \frac{e^{i2\pi\left(2^{n+\delta}\theta - y'/2^{n+\delta}\right)} - 1}{e^{i2\pi\left(\theta - y'/2^{n+\delta}\right)} - 1} \right|^2$ can be given.

Let $a = \left| e^{i2\pi\left(2^{n+\delta}\theta - y'\right)} - 1 \right| = \left| e^{i2\pi\left(2^{n+\delta}\varepsilon\right)} - 1 \right|$, and $b = \left| e^{i2\pi\left(\theta - y'/2^{n+\delta}\right)} - 1 \right| = \left| e^{i2\pi(\varepsilon)} - 1 \right|$, so that

$$p_{y'} = \frac{1}{2^{2n}} \frac{a^2}{b^2} = \frac{1}{2^{2n}} \frac{\left| e^{i2\pi\left(2^{n+\delta}\varepsilon\right)} - 1 \right|^2}{\left| e^{i2\pi(\varepsilon)} - 1 \right|^2}. \tag{31}$$

To express the lower bound on probability $p_{y'}$ in the proposed Quantum-SVD approach, we need a *lower bound* on $a$ and an *upper bound* on $b$. The ratio of the *minor arc length* to the *chord length* is at most $\pi/2$, since $\frac{2\pi|\varepsilon|2^{n+\delta}}{a} \leq \frac{\pi}{2}$, which implies[1,2,6,8]

$$a \geq 4|\varepsilon|2^{n+\delta}. \tag{32}$$

In the classical QFT algorithm, $\frac{2\pi|\varepsilon|2^{n}}{a} \leq \frac{\pi}{2}$, and $a \geq 4|\varepsilon|2^{n}$, hence

$$p_{y'} = \frac{1}{2^{2n}} \frac{a^2}{b^2} = \frac{1}{2^{2n}} \frac{\left| e^{i2\pi(2^{n}\varepsilon)} - 1 \right|^2}{\left| e^{i2\pi(\varepsilon)} - 1 \right|^2}. \tag{33}$$

In Fig. 7. we illustrated the *lower* bound of parameter *'a'* of classical QFT and our Quantum-SVD approach.

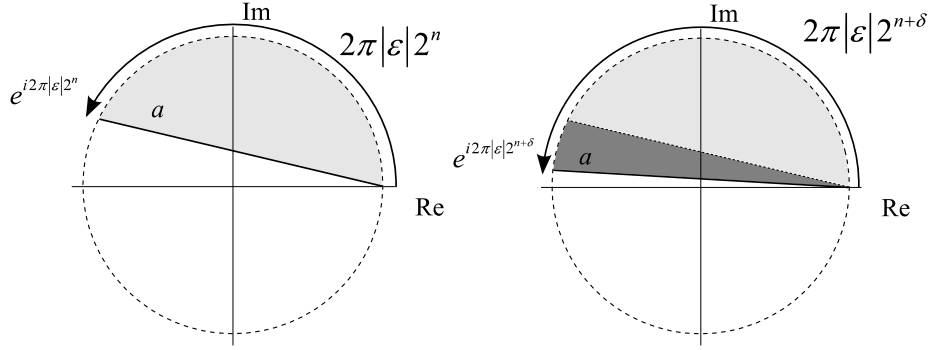

**Fig. 7.** Comparison of lower bounds on parameter '$a$' of classical QFT and quantum-SVD algorithms.

Similarly, $b = \left|e^{i2\pi(\varepsilon)} - 1\right|$ and assuming that the ratio of arc length to chord is at least 1 leads to $\dfrac{2\pi|\varepsilon|}{b} \geq 1$. The *upper bound* for $b$ is[1,2,8]

$$b \leq 2\pi|\varepsilon|. \tag{34}$$

From the bounds for $a$ and $b$ the following result follows:

$$p_{y'} \geq \frac{1}{2^{2n}} \frac{\left|4|\varepsilon|2^{n+\delta}\right|^2}{\left|2\pi|\varepsilon|\right|^2} = \frac{1}{2^{2n}} \frac{16|\varepsilon|^2 \, 2^{2n} \cdot 2^{2\delta}}{4\pi^2 |\varepsilon|^2}$$
$$= \frac{4 \cdot 4^{\delta}}{\pi^2} > 0.4 \cdot 4^{\delta} \approx 0.4 \cdot 1.4 \approx 0.56. \tag{35}$$

As classical QFT transformation, in the quantum-SVD approach we have $b \leq 2\pi|\varepsilon|$, hence we obtain probability

$$p_{y'} \geq \frac{1}{2^{2n}} \frac{\left|4|\varepsilon|2^n\right|^2}{\left|2\pi|\varepsilon|\right|^2} = \frac{1}{2^{2n}} \frac{16|\varepsilon|^2 \, 2^{2n}}{4\pi^2|\varepsilon|^2} = \frac{4}{\pi^2} > 0.4. \tag{36}$$

In Fig. 8. we illustrated the geometrical picture of classical QFT and our Quantum-SVD algorithm on the *upper* bound of parameter *'b'*.

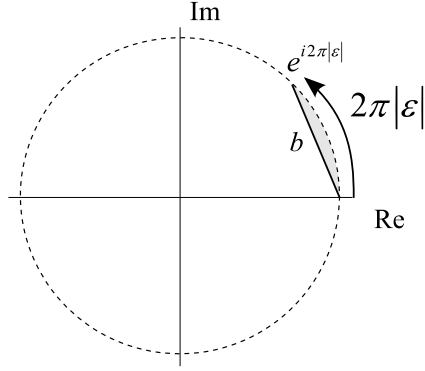

**Fig. 8.** The upper bound on parameter '$b$' is the same for classical QFT and quantum-SVD algorithms.

As we can conclude, using the classical QFT algorithm, the probability that every single one of the bits is correct is $p_{y'} > 0.4$. In the proposed Quantum-SVD approach, the probability, that every single one of the bits is correct is

$$p_{y'} > 0.4 \cdot 4^{\delta}. \tag{37}$$

The upper bound on the probability of obtaining *inaccurate results* in the approximation of the QFT algorithm could be given in the same way. For a given value of $y'$ we have $e^{2\pi i \theta} = e^{2\pi i \left(\frac{y'}{2^{n+\delta}} + \varepsilon\right)}$ for error threshold $\frac{\gamma}{2^{n+\delta}} \leq |\varepsilon| < \frac{1}{2}$, where $\gamma$ is an arbitrary positive number. We have $p_{y'} = \frac{1}{2^{2n}} \frac{a^2}{b^2} = \frac{1}{2^{2n}} \frac{\left|e^{i2\pi\left(2^{n+\delta}\varepsilon\right)} - 1\right|^2}{\left|e^{i2\pi(\varepsilon)} - 1\right|^2}$, but we use the trivial fact that $a \leq 2$ and $b \geq 4|\varepsilon|$. The upper bound on $p_{y'}$ is $p_{y'} \leq \frac{1}{2^{2n}} \frac{a^2}{b^2} = \frac{1}{2^{2n}} \frac{|2|^2}{\left(4|\varepsilon|\right)^2} = \frac{1}{2^{2n}} \frac{4}{\left(4|\varepsilon|\right)^2} = \frac{1}{4\gamma^2}$. If our assumption is, that the *error threshold* is only $|\varepsilon| \geq 2^{-(n-\delta)}$, then

$$p_{y'} \leq \frac{1}{2^{2n}} \frac{4}{\left(4\left|2^{-(n-\delta)}\right|\right)^2} = \frac{1}{2^{2n}} \frac{4}{\left(16 \cdot 2^{-2n} \cdot 2^{2\delta}\right)} = \frac{1}{4 \cdot 4^{\delta}} \approx \frac{1}{4 \cdot 1.4} \approx 0.178, \tag{38}$$

thus $\gamma = \sqrt{4^\delta} \approx \sqrt{1.4}$. It means that the probability of obtaining a highly inaccurate result is at most

$$\frac{1}{4 \cdot 4^\delta} = \frac{1}{4 \cdot 1.4} \approx 0.178. \tag{39}$$

Implying a larger bound on the error threshold $|\varepsilon|$ results in less accurate approximations, and the probability of obtaining the corresponding value of $y$ becomes very small. The probability of *correctness* converges quickly to 1 exponentially as the procedure is repeated.

In the classical QFT algorithm, we can use the same methods to put upper bounds on the probability of obtaining inaccurate results, since for a given value of $e^{2\pi i \theta} = e^{2\pi i \left(\frac{y'}{2^n} + \varepsilon\right)}$ for error threshold $\frac{\gamma}{2^n} \leq |\varepsilon| < \frac{1}{2}$, we get

$$p_{y'} \leq \frac{1}{2^{2n}} \frac{4}{\left(4|2^{-n}|\right)^2} = \frac{1}{2^{2n}} \frac{4}{16\left(2^{-2n}\right)} = \frac{1}{4} = 0.25. \tag{40}$$

Hence, if we assume, that $|\varepsilon| \geq 2^{-n}$, the probability of obtaining the corresponding value of $y'$ is at most 0.25, in the classical QFT algorithm.

In Fig. 9. we compared the ratio of parameters, '*a*' and '*b*', for classical QFT, and the proposed Quantum-SVD approach.

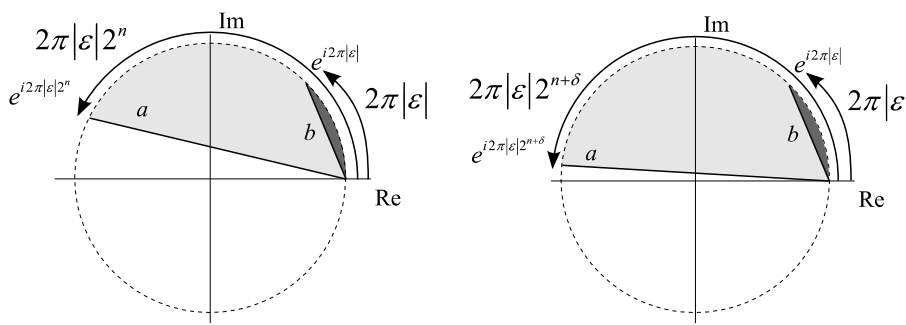

**Fig. 9.** Comparison of parameters '*a*' and '*b*' for ordinary QFT and Quantum-SVD approach.

In Fig. 10. we compared the approximation errors of QFT and Quantum-SVD approaches for non-uniform angles.

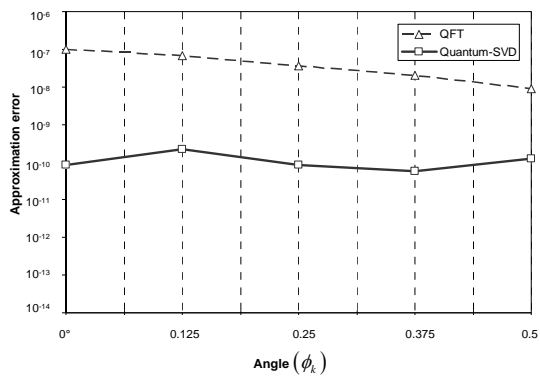

**Fig. 10.** Approximation errors of non-uniform QFT exponentials by ordinary QFT and Quantum-SVD approach.

The error of the Quantum-SVD approach is three orders of magnitude lower than the approximation error of ordinary QFT method.

## 6  Conclusions

The presented Quantum-SVD algorithm is a novel approach for the computation of the Quantum Fourier Transformation of non-uniformly distributed quantum states. Comparing the error probabilities for generic matrix with non-uniform angles, the error given by the Quantum-SVD approach is some orders lower than error given by interpolation approach. The proposed Quantum-SVD algorithm is based on singular value decomposition, and has the same complexity as the standard Quantum Fourier Transform. While, the Quantum-SVD based approach has a higher computational cost for the calculation of the projection components, these costs simply can be negated by using the power of quantum computation.